\documentclass{article} 
\usepackage{iclr2025_conference,times}

\usepackage{amsmath,amsfonts,bm}

\def\eqref#1{equation~\ref{#1}}

\def\1{\bm{1}}

\DeclareMathAlphabet{\mathsfit}{\encodingdefault}{\sfdefault}{m}{sl}
\SetMathAlphabet{\mathsfit}{bold}{\encodingdefault}{\sfdefault}{bx}{n}

\usepackage{natbib}
\usepackage{hyperref}
\usepackage{url}
\usepackage{graphicx}
\usepackage{fvextra}

\title{Leveraging LLMs for Top-Down Sector \\Allocation in Automated Trading}

\author{\textbf{Ryan Quek Wei Heng}$^{1}$ \quad \textbf{Edoardo Vittori}$^{2}$ \quad \textbf{Keane Ong}$^{1,3}$ \quad \textbf{Rui Mao}$^{4}$ \\
    \textbf{Erik Cambria}$^{4}$ \quad \textbf{Gianmarco Mengaldo}$^{1,3,5,6}$ \\
    $^{1}$College of Design and Engineering, National University of Singapore \\
    $^{2}$CVA Management and A.I. Investments, IMI Corporate and Investment Banking, Intesa Sanpaolo \\
    $^{3}$Asian Institute of Digital Finance, National University of Singapore \\
    $^{4}$College of Computing and Data Science, Nanyang Technological University \\
    $^{5}$Sustainable and Green Finance Institute, National University of Singapore \\
    $^{6}$Honorary Research Fellow, Imperial College London \\
    \texttt{ryanquekweiheng@u.nus.edu} \\
    \texttt{edoardo.vittori@intesasanpaolo.com} \\
    \texttt{keane.ongweiyang@u.nus.edu} \\
    \texttt{rui.mao@ntu.edu.sg} \\
    \texttt{cambria@ntu.edu.sg} \\
    \texttt{mpegim@nus.edu.sg}
}

\iclrfinalcopy 

\begin{document}

\maketitle
\noindent
\begin{flushleft}
\textit{Preprint accepted to ICLR Workshop Advances in Financial AI: Opportunities, Innovations, and Responsible AI on March 5, 2025}
\end{flushleft}

\begin{abstract}
This paper introduces a methodology leveraging Large Language Models (LLMs) for sector-level portfolio allocation through systematic analysis of macroeconomic conditions and market sentiment. Our framework emphasizes top-down sector allocation by processing multiple data streams simultaneously, including policy documents, economic indicators, and sentiment patterns. Empirical results demonstrate superior risk-adjusted returns compared to traditional cross-momentum strategies, achieving a Sharpe ratio of 2.51 and portfolio return of 8.79\% versus -0.61 and -1.39\% respectively. These results suggest that LLM-based systematic macro analysis presents a viable approach for enhancing automated portfolio allocation decisions at the sector level.
\end{abstract}

\section{Introduction}
The rapid advancement of Large Language Models (LLMs) has catalyzed significant innovations in algorithmic trading and investment management, with numerous studies exploring their potential for stock selection and portfolio optimization~\citep{ding2024large}. 
However, while existing research has predominantly focused on bottom-up security selection using company-specific metrics (see e.g., \cite{yu2024finmem, zhang2024finagent, wang2024llmfactor}), there remains a crucial gap in understanding how LLMs can enhance top-down investment strategies, particularly in the context of sector allocation within a portfolio.

Macroeconomic factors fundamentally shape investment decision-making, influencing stock valuations through both economic channels and market sentiment~\citep{jareno2016us, jabeen2022empirical, ma2024quantitative}. 
A top-down investment approach, based on macroeconomic analysis, enables investors to anticipate broad market trends, optimize portfolio allocation across economic cycles, and identify systematic risks before they manifest in individual equity prices. 
These macroeconomic forces generate heterogeneous sectoral responses; for instance, inflationary pressures typically constrain manufacturing growth while benefiting services and agricultural sectors~\citep{chaudhry2013does}. 
Moreover, hawkish Federal Reserve communications not only elevate borrowing costs and alter corporate decision-making but also intensify investor concerns about financially vulnerable firms, amplifying negative sentiment~\citep{cieslak2023tough,xinfina}. 

This sentiment-driven reaction often precipitates market movements that exceed what fundamentals alone would otherwise suggest. 
Key macroeconomic variables—including monetary policy, inflation dynamics, and economic growth trajectories—propagate through the financial system, influencing both firms' cost of capital and aggregate consumption patterns.
Given that the task of constructing a portfolio based on sector allocation is one that requires piecing together information from different sources, this paper seeks to address the following \textbf{Research Question} -- How can LLMs be leveraged to enhance top-down sector allocation strategies by integrating macroeconomic analysis with market sentiment for automated portfolio construction?

To respond to this question, we introduce a novel LLM-based methodology that emphasizes top-down sector allocation through integrated analysis of macroeconomic and market sentiment data. 
While existing approaches incorporate various data sources including sentiment and fundamentals, they typically focus on bottom-up analysis of individual securities. 
Our framework leverages LLMs to systematically process and synthesize multiple data streams simultaneously (including policy documents, economic indicators, and sentiment patterns), enabling dynamic adjustment of sector allocations based on market conditions. 
By automating the extraction and interpretation of these macro financial relationships, our framework enhances the responsiveness of sector allocation strategies through a predominantly top-down lens. 
This provides a more systematic approach to capturing sentiment-driven price movements, offering new insights into sector-level portfolio construction that complement traditional security-level analysis.

Results obtained through backtesting indicate that our sector allocation strategy significantly outperforms traditional cross-momentum approaches. 
Over the testing period, our methodology achieved a positive return of 8.79\% and a Sharpe ratio of 2.51, compared to a loss of 1.39\% and a Sharpe ratio of -0.61 for the cross-momentum strategy. 
These results suggest that our top-down LLM-based approach effectively captures sector-level opportunities while maintaining strong risk-adjusted performance, demonstrating the potential value of incorporating systematic macro and sentiment analysis in portfolio allocation decisions.

\section{Related Works}
In this section, we cover related literature on the following four topics: cross-sectional momentum investment, top-down investment, usage of LLMs as trading agents, and finance specific Aspect based Sentiment Analysis.

\subsection{Cross-sectional momentum investment}
\label{sec : cross-sectional-momentum}
Traditional cross-sectional momentum strategies rank assets based on their relative performance, taking long positions in top performers and short positions in underperformers. The classical approach introduced by~\citet{jegadeesh1993returns} ranks stocks based on past returns, while modern variations incorporate volatility normalization and machine learning models trained to minimize mean squared error before ranking assets to construct portfolios.

Learning to Rank (LTR) algorithms further improve upon traditional approaches by explicitly optimizing for ranking quality rather than prediction accuracy~\citep{poh2020building}. These algorithms, which include pairwise methods like RankNet~\citep{burges2005learning} and LambdaMART~\citep{wu2010adapting} and listwise methods like ListNet~\citep{cao2007learning}, learn the relative ordering between instruments directly. When applied to cross-sectional momentum strategies, LTR methods demonstrate significant improvements over traditional approaches, achieving approximately three times higher Sharpe ratios.

\subsection{Top-down investment}
\label{sec: macro-driven investment}
The relationship between macroeconomic indicators and stock markets has been widely studied, demonstrating their significant impact on equity values.

Early research on Sri Lanka’s stock market~\citep{gunasekarage2004macroeconomic} found that interest rates (measured by the treasury bill rate) and inflation (measured by the consumer price index) strongly influence stock prices, with the treasury bill rate exerting the most significant effect over a 17-year period (1985–2001).

More recent studies leverage macroeconomic regime modeling to enhance equity factor investing strategies~\citep{nuriyev2024augmenting}. Techniques such as Hidden Markov Models and meta-correlation-based clustering have outperformed traditional strategies by dynamically adjusting portfolio weights based on economic conditions, favoring quality factors during crises and momentum factors during growth phases. This further underscores the critical role of macroeconomic indicators in shaping stock market behavior.

\subsection{LLM Trading Agents}
Previous works~\citep{ding2024large} on LLM-based trading agents can be broadly categorized into (i) LLM as a Trader and (ii) LLM as an Alpha Miner. This literature review focuses on the former, examining works in News-Driven and Reflection-Driven LLMs as Traders.

\subsubsection{LLM as a Trader - News Driven}
\label{sec: LLM agent-New Driven}
News-driven LLMs integrate stock news and macroeconomic updates into their prompt context before instructing the LLMs to predict stock price movements.

Studies~\citep{lopez2023can, wu2024portfolio} show that both proprietary and open-source LLMs, even without specialized financial training, can effectively analyze news sentiment for stock market prediction. Backtesting a simple long-short strategy based on sentiment scores from news headlines has demonstrated promising results.

A more advanced approach~\cite{fatouros2024can} enhances the approach by summarizing and refining news data while reasoning about its relationship with stock price movements. A memory module stores these summaries, which are later retrieved as contextual "recommendations" for trading decisions. By leveraging progressive daily news summaries, macroeconomic insights, and stock price momentum summaries, the MS-Top10-Cap-GPT strategy outperformed the S\&P 100 index by approximately 30\%.

\subsubsection{LLM as a Trader - Reflection Driven}
\label{sec: LLM Agent-Reflection}
Apart from memories, reflections can be used in LLM decision making. In this context, Reflections~\citep{park2023generative} are defined as high-level knowledge and insights progressively aggregated from raw memories and observations. The inclusion of memory and reflection in LLM-based algorithms offers significant benefits such as mitigating the risk of hallucinations~\citep{ji2023towards} and obtaining high-level understanding of the environment~\citep{park2023generative}.

FinMem~\citep{yu2024finmem} introduces a trading agent with layered memorization and characteristics. The raw inputs, such as daily news and financial reports, are summarized into memories. Upon the arrival of new observations, the relevant memories are retrieved and integrated with these observations to produce reflections. Both memories and reflections are stored in a layered memory bucket. During the trading phase, these memories and reflections are retrieved and utilized by the decision-making module to generate the final trading decisions. The retrieval method considers the recency, relevancy, and importance of the information.

Similarly, FinAgent~\citep{zhang2024finagent} extends this approach with a multimodal module that processes numeric, text, and image data. It enhances decision making by incorporating technical indicators like MACD and RSI, along with analyst guidance, demonstrating superior backtesting performance over FinMem.

\subsection{Finance specific Aspect based sentiment analysis}
\label{sec: ABSA}
Traditional sentiment analysis in finance has historically focused on broad positive/negative classifications of text towards stocks, but markets and financial decisions require more nuanced understanding. Aspect based Sentiment Analysis (ABSA) has emerged as a crucial advancement by enabling the extraction and analysis of sentiment toward specific aspects or features within financial text~\cite{du2024financialsentiment,malpub,maamul}. Previous works of finance-specific ABSA~\citep{ong2023finxabsa} highlight the separation of sentiment across different financial aspects for explainable sentiment, allowing for assessment of sentiment towards specific risk factors or market aspects. 

On the other hand, LLMs without finetuning were used for the task on finance specific ABSA. Through the Heterogeneous multi-agent Discussion (HAD) framework~\citep{xing2024designing}, the accuracy and F1 scores obtained on established datasets like FiQA~\citep{maia201818}, demonstrate the potential of LLMs to perform finance specific ABSA without specific financial training.

\section{Data Source}
This section provides an overview of the data sources utilized for this project. The backtesting period was limited to January 2019 through June 2019 due to constraints on computational resources and cost.

Our backtesting framework focused exclusively on S\&P 500 constituents to maintain a well-defined set of liquid, large-cap stocks. To mitigate survivorship bias, we dynamically updated our universe to reflect real-time changes in the index composition. For instance, when First Republic Bank (FRC) replaced SCANA Corporation (SCG) following its acquisition by Dominion Energy on 2nd January 2019~\citep{SPGlobal2018}, our set of stock used for backtesting was adjusted accordingly. Historical market data, including OHLC (open, high, low, close) prices and trading volumes, was sourced from the~\citet{AlphaVantage} API.

The following Macro data were selected to be representative of the macroeconomic variables-interest rates, inflation, employment and economic growth-that represent the distinct economic categories as outlined by~\citet{nuriyev2024augmenting}.

\begin{itemize}
    \item Inflation:
    \begin{itemize}
        \item Consumer Price Index (CPI)~\citep{BLS_CPI}
        \item Producer Price Index (PPI)~\citep{BLS_PPI}
        \item Personal Consumption Expenditures (PCE)~\citep{BEA_PCE}
    \end{itemize}
    \item Employment:
    \begin{itemize}
        \item Non-Farm Payrolls (NFP)~\citep{BLS_PAYEMS}
    \end{itemize}
    \item Economic Growth:
    \begin{itemize}
        \item Purchasing Managers' Index (PMI)~\citep{Investing_ISM_PMI}
    \end{itemize}
    \item Interest Rates:
    \begin{itemize}
        \item Federal Open Market Committee (FOMC) minutes~\citep{FOMC_Calendar}
    \end{itemize}
\end{itemize}

Notably, the PCE data for 2018-Dec, 2019-Jan and 2019-Feb were released off schedule due to the partial federal government shutdown. As such, the latest available data from previous months was used instead.

By using keywords such as stock-specific company names and global market identifiers, news articles were scraped from the web via~\citet{NewsAPI} to be used as sentiment data. A total of 300000 articles were used to form the backtesting window of 6 months, 2019-Jan to 2019-Jun. Each news article consisted of the published date, title, description and content.

\section{Design Process}
\begin{figure}
    \centering
    \includegraphics[width=0.92\linewidth]{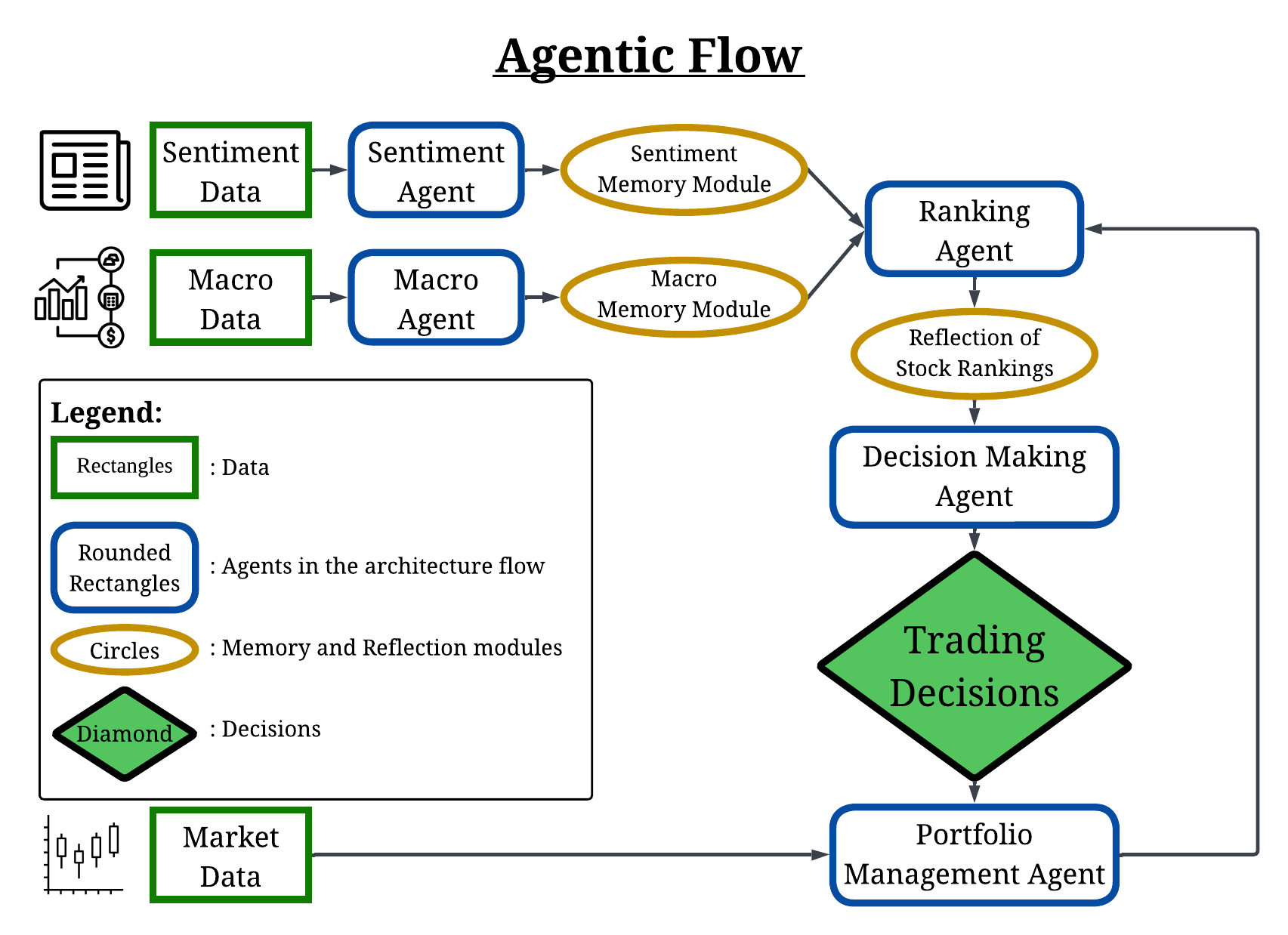}
    \caption{Overview of agentic flow designed. Icons obtained from Flaticon.com}
    \label{fig: agentic_flow_architecture}
\end{figure}

Top-down sector allocation investing requires a comprehensive understanding of both macroeconomic conditions and company-specific sentiments. As shown in our architecture diagram (refer to Fig~\ref{fig: agentic_flow_architecture}), we address this through a dual-stream processing pipeline that analyzes news articles and macroeconomic data in parallel. The system integrates these inputs through a Ranking Agent which generates stock rankings that inform sector-level investment decisions. The architecture incorporates memory and reflection modules (represented as circles in the diagram) as discussed in Section~\ref{sec: LLM Agent-Reflection} to help the system develop a high-level understanding of the macroeconomic environment.

$ $

As outlined in Section~\ref{sec: ABSA}, recent research has demonstrated that LLMs can perform financial sentiment analysis without domain-specific training, guiding our design of the sentiment analysis pipeline. 
The Sentiment Agent, powered by the DeepSeek-7B-Chat model~\citep{DeepSeekLLM7BChat}, processes news articles through two critical analyses: Named Entity Recognition (NER) and ABSA. NER is essential because our keyword search encompasses both stock-specific and broader market-relevant terms, enabling the identification of articles pertinent to target stocks that might be missed by simple keyword matching. The ABSA component analyzes article descriptions and content to identify multiple aspects of sentiment, providing the Ranking Agent with granular information about how macroeconomic conditions may amplify or diminish stock-specific sentiments through various channels. 
The resulting sentiment data (stock ticker and aspect sentiment pairs) is maintained in a dedicated Sentiment Memory Module, which serves as a persistent repository for historical sentiment information.
The DeepSeek-Chat-7B model, selected for three key advantages: its open-source nature, efficient inference speed, and cost-effective operation. The Sentiment Agent executes the analysis through a specialized prompt template (detailed in Appendix, Section~\ref{sec: sentiment-prompt}). 


Parallel to the sentiment pipeline, the macroeconomic data stream is processed by a Macro Agent. Since raw macroeconomic values lack contextual meaning for LLMs, this component calculates month-over-month percentage changes for key economic indicators including CPI, PPI, PCE, NFP, and PMI. These relative changes provide meaningful representations of macroeconomic trends that LLMs can effectively interpret and analyze. Additionally, the Macro Agent processes FOMC meeting minutes using DeepSeek-Chat-7B to generate concise summaries to be stored as memories in the Macro Memory Module for subsequent retrieval. For more information about the prompts used, please refer to the Appendix, Section~\ref{sec : summarize-fomc-prompt}.

The core of the agentic architecture is the Ranking agent, which retrieves memories from the News and Macro modules, as well as current portfolio positions from the Portfolio Management Agent, to generate a Stock Ranking reflection for a single day. To prevent look-ahead bias, the agent operates under strict temporal constraints, where it only processes news memories published on the previous trading day and macro memories based on their official release dates, using the latest available data points for each economic indicator. This temporal filtering mechanism ensures that all trading decisions are made using only information that would have been available to market participants at the time of trading, maintaining the practical applicability of our backtesting results.

The Ranking Agent implements a hierarchical top-down sector allocation strategy through three sequential analytical phases: (i) macroeconomic analysis, (ii) sector allocation, and (iii) sentiment integration. In the initial phase, the agent analyzes inflation trends, evaluates economic strength indicators, and assess monetary policy trajectories. This macroeconomic assessment helps identify helps which sectors perform the best in the given macroeconomic environment. The second phase focuses on portfolio optimization, where the agent determines optimal allocations across the 11 GICS sectors~\citep{GICS_mapbook_brochure} and evaluates necessary portfolio adjustments by comparing target allocations against current sector exposures. In the final phase, leverages sentiment-aspect pairs from the Sentiment Module to select stocks within each targeted sector. This refinement process evaluates how prevailing macroeconomic conditions may amplify positive sentiments or mitigate negative ones for specific business aspects, enabling the selection of stocks best positioned for both long and short positions within their respective sectors. More information on the prompt used to generate the stock ranking reflections is available in Appendix, Section~\ref{sec : prompt-ranking-agent-top-down}.

We employ the Llama-3.3-70B-Instruct model~\citep{meta2024llama} as the foundation for our Ranking Agent. This model selection was motivated by several key factors: its open-source nature, its extensive training on instruction-following tasks using publicly available datasets and over 25M synthetically generated examples. These characteristics make it particularly well-suited for interpreting and executing the complex, multi-phase analysis required for our sector allocation strategy.
The daily stock ranking reflection is processed by a Decision-making Agent, implemented using Llama-3.2-3B~\citep{meta2024llama3_2_3b}, which converts qualitative rankings into structured JSON trading decisions. Given the standardized format of the reflections generated by the Ranking Agent, we opted for a lighter-weight instruction-following model to efficiently parse and transform these reflections into actionable trade signals. These decisions are then routed to the Portfolio Management Agent, which interfaces with real-time market data to optimize trade execution.

The Portfolio Management Agent implements a conservative capital management strategy designed to balance portfolio performance with risk mitigation. To maintain sufficient liquidity for opportunistic trading while preventing overexposure, the agent enforces a 90\% maximum capital utilization threshold. During the execution of trades, new positions are initiated only when they do not create conflicts with existing holdings, whereas positions are fully liquidated when new rankings indicate a shift in stock sentiment outlook. Furthermore, to maintain alignment with our investment scope, positions in stocks removed from the S\&P 500 index during the backtesting window are automatically liquidated.


\section{Methodology}
To evaluate the effectiveness of this architecture, we implement two distinct methodologies: our proposed top-down sector-based allocation strategy that leverages on LLM-driven macroeconomic analysis, and a baseline cross-momentum strategy that build a portfolio purely based on momentum factors. 
Compared to the top-down sector allocation strategy, the ranking agent in the cross-momentum strategy would only carry out step (iii), integrating sentiment and macroeconomic information, to generate the reflection of stock rankings. Also, the size of the position of each trade is fixed at 1\% of the portfolio. More information is available in Section~\ref{sec : prompt-ranking-agent-cross-sectional} of the Appendix.

An initial capital of USD100M was used to simulate real-world institutional trading conditions and provide sufficient scale to meaningfully assess the impact of transaction costs on strategy performance.
Trading costs were tracked through a comprehensive framework that accounts for both explicit and implicit costs. A commission rate of 10 basis points per trade was applied to reflect standard institutional brokerage fees, while market impact costs of 10 basis points were included to account for price slippage during execution. 


\begin{table}[h]
\caption{Backtesting results from Jan-2019 to Jun-2019}
\label{results-table}
\begin{center}
\begin{tabular}{lcc}
\multicolumn{1}{c}{\bf Strategy}  &\multicolumn{1}{c}{\bf PCT Change in Portfolio} &\multicolumn{1}{c}{\bf Sharpe Ratio}
\\ \hline \\
Cross-Momentum             &-1.39\% & -0.61 \\
Sector-Allocation       &\textbf{8.79\%} & \textbf{2.51} \\
\end{tabular}
\end{center}
\end{table}

\section{Results}
The portfolio percentage change and Sharpe ratio were selected as the primary metrics for evaluating the strategy, as they provide complementary perspectives on both performance and risk-adjusted returns. Portfolio percentage change offers a normalized view of performance, allowing for direct comparisons across different portfolio sizes and time periods by expressing gains and losses as a proportion of the initial investment. 

This facilitates a standardized assessment of the strategy’s performance. In contrast, the Sharpe ratio, which gauges excess returns relative to volatility, was chosen to evaluate the risk-adjusted efficiency of the strategy. Together, these metrics offer a comprehensive evaluation of the strategy’s profitability and its ability to generate sustainable returns.

The results presented in the Table~\ref{results-table} are calculated as follows: Portfolio percentage change is derived from the final portfolio value, which includes unrealized profits, minus trading costs. The Sharpe ratio is computed by dividing the daily excess returns by the standard deviation of the portfolio's returns, providing a measure of risk-adjusted performance.

The results indicate that the Sector-Allocation strategy significantly outperformed the Cross-Momentum approach across both metrics. While the Cross-Momentum strategy resulted in a negative portfolio return of -1.39\% and a Sharpe ratio of -0.61, suggesting poor risk-adjusted performance, the Sector-Allocation strategy delivered superior results with an 8.79\% portfolio gain and a strong Sharpe ratio of 2.51. The positive Sharpe ratio for the Sector-Allocation strategy indicates that it not only generated better absolute returns but also achieved this performance with favorable risk-adjusted characteristics. This suggests that the sector-based approach was more effective at capturing market opportunities while managing volatility. The stark contrast in performance metrics between the two strategies highlights the benefits of incorporating top-down sector allocation in trading strategies.

\section{Limitations and Future Directions} 
While our results demonstrate promising performance, several limitations and opportunities for future research warrant discussion. A primary limitation is the relatively short backtesting period from 2019 to 2024, constrained by computational resources and costs. Extending this period would provide more robust validation across different market cycles and economic conditions. Additionally, our current implementation could benefit from larger language models, which have demonstrated superior capabilities in complex Natural Language Processing tasks. Specifically, using models with more parameters could enhance the accuracy of ABSA, NER, and the interpretation of FOMC minutes by capturing more subtle linguistic nuances and relationships.

Several promising directions exist for future research. First, expanding the model's data inputs to include company fundamental data, particularly quarterly earnings reports, could reveal whether incorporating bottom-up financial metrics alongside our top-down approach yields meaningful improvements in portfolio performance~\citep{ong2025survey}. 

Second, the ranking mechanism could be enhanced through the implementation of LTR algorithms as mentioned in Section~\ref{sec : cross-sectional-momentum} or Reinforcement Learning techniques. These approaches have demonstrated success in improving cross-sectional strategies, suggesting potential benefits for sector allocation. By incorporating historical ranking performance as feedback, these methods could dynamically optimize allocation decisions and enhance the model's adaptability to changing market conditions.

Additionally, integrating FINMEM's~\citep{yu2024finmem} architecture could enhance the model's ability to process multi-timeframe financial data while maintaining important information over extended periods, potentially capturing longer-term patterns and relationships that human traders might overlook.

Lastly, financial Explainable Artificial Intelligence techniques~\citep{yeo2023comprehensive} such as counterfactual explanations, can be used to evaluate performance in alternative scenarios. These methods provide insight into LLM decision making and support regulatory compliance.

\section{Conclusion}
In this paper, we introduced a novel methodology leveraging Large Language Models for top-down sector allocation through integrated analysis of macroeconomic and market sentiment data. Our approach demonstrates that LLMs can effectively process and synthesize multiple data streams simultaneously, enabling dynamic sector allocation adjustments based on market conditions. The empirical results validate the effectiveness of this approach, with our strategy achieving an 8.79\% return and a Sharpe ratio of 2.51, significantly outperforming traditional cross-momentum approaches which recorded a -1.39\% return and -0.61 Sharpe ratio during the same period.

These results suggest that systematic top-down analysis using LLMs offers a viable approach for enhancing portfolio allocation decisions. The framework's ability to process unstructured data and identify macro financial relationships represents a step forward in automating complex market analysis tasks. As the field of quantitative investing continues to evolve, our methodology demonstrates the potential for LLMs to complement existing investment strategies by providing a more systematic approach to sector-level portfolio construction.

\subsubsection*{Acknowledgments}
The authors gratefully acknowledge the support of the MOE Tier 1 Startup project, under grant number \#22-3565-A0001-1.
The research is also supported by MOE Academic Research Fund Tier 2 (STEM RIE2025 Award MOE-T2EP20123-0005) and RIE2025 Industry Alignment Fund -- Industry Collaboration Projects (IAF-ICP) (Award I2301E0026), administered by A*STAR, as well as supported by Alibaba Group and NTU Singapore.

\bibliography{iclr2025_conference}

\bibliographystyle{plainnat}

\section{Appendix}
\subsection{Prompt for Sentiment pipeline}
\label{sec: sentiment-prompt}
\begin{Verbatim}[breaklines=true, breakanywhere=true]
    You are a financial sentiment analyzer. Your task is to analyze news articles about companies and extract sentiment information about different aspects of the company mentioned in the article. Respond ONLY with a JSON object, no additional text or markdown. 

        Instructions:
        1. Analyze the provided news article's title, description, and content.
        2. Identify the main stock/company being discussed.
        3. Extract 3 to 5 key aspects discussed in the article (e.g., earnings, revenue, management, products, market position, growth, competition).
        4. For each aspect, determine the sentiment on a scale:
        - positive (1)
        - neutral (0)
        - negative (-1)
        5. Return the analysis in the following JSON format exactly, replacing the example values with your analysis.
        
        {"stock": "AAPL",
        "aspect_sentiment_pairs": [
            ["revenue", 1],
            ["product_performance", -1],
            ["services", 1]
        ]}


        Rules:
        - Only include information that is explicitly discussed in the article
        - Only include aspects that belong to the list of relevant aspects to look out for
        - Base sentiment strictly on the article's content, not external knowledge
        - Be consistent with aspect naming (e.g., always use "revenue" instead of mixing "revenue" and "sales")
        - Don't include duplicate aspects
        - Limit to the most significant 3-5 aspects mentioned
        - Use the most commonly known stock ticker
        - If no clear stock ticker is mentioned, use the company name in the stock field, 
        

        List of relevant aspects to look for:
        - revenue/sales
        - earnings/profit
        - market_share
        - product_performance
        - management
        - growth
        - competition
        - regulatory
        - innovation
        - customer_demand
        - operational_efficiency
        - partnerships
        - risk
        - strategy

        Example Analysis:

        Input:
        Title: EGG Reports Record Q4 Revenue Despite Product Sales Miss
        Description: EGG posts strong services growth but flagship product disappoints
        Content: EGG Inc. reported its highest-ever fourth-quarter revenue of $89.5 billion, though Product sales fell short of analyst expectations. The company's services division saw remarkable growth, up 16% year-over-year, helping offset the weaker hardware performance. CEO Tim Cook expressed confidence in the company's product pipeline but acknowledged supply chain challenges.

        Output:
        {"stock": "EGG",
        "aspect_sentiment_pairs": [
            ["revenue", 1],
            ["product_performance", -1],
            ["services", 1],
            ["supply_chain", -1]
        ]}
        End of example"""
\end{Verbatim}

\subsection{Prompt for summarizing FOMC minutes}
\label{sec : summarize-fomc-prompt}
\begin{Verbatim}[breaklines=true, breakanywhere=true]
    Please analyze the following FOMC meeting minutes and provide a structured analysis focusing on these key aspects:

        1. Interest Rate Policy and Outlook
        - Identify explicit statements about current interest rate decisions
        - Extract any forward guidance or projections about future rate movements
        - Note any dissenting views or alternative scenarios discussed

        2. Economic Assessment
        - Summarize the Committee's view on:
        * GDP growth and economic activity
        * Labor market conditions
        * Inflation rates and price stability
        * Financial market conditions

        3. Risk Analysis
        - List major risks to the economic outlook
        - Detail both upside and downside risks

        Meeting Minutes:
        {text}

        Provide analysis in a clear, structured format.
\end{Verbatim}

\subsection{Prompt for Ranking Agent in top-down strategy}
\label{sec : prompt-ranking-agent-top-down}
\begin{Verbatim}[breaklines=true, breakanywhere=true]
You are a quantitative macro strategist specializing in top-down allocation strategies. Your task is to analyze macroeconomic conditions first, then use sentiment data to select stocks that align with the macro outlook.

        Given the following inputs:
        [MACRO DATA TRENDS]
        - Latest trend readings: CPI {value}, PPI {value}, PCE {value}, NFP {value}, PMI {value}
        
        [FOMC MINUTES SUMMARY]
        - Recent FOMC minutes summary: {key_points}

        [STOCK UNIVERSE]
        - List of S&P 500 stocks with their sentiment data
        - Format: date|ticker|aspect_sentiment_pairs
            
        [CURRENT PORTFOLIO]
        - List of current positions:
            - Long positions: [(TICKER, weight)]
            - Short positions: [(TICKER, weight)]

        For sector analysis, use the 11 GICS sectors:
        - Information Technology
        - Financials
        - Healthcare
        - Consumer Discretionary
        - Consumer Staples
        - Industrials
        - Energy
        - Materials
        - Communication Services
        - Utilities
        - Real Estate

        Analysis Process:
        1. Macro Environment Assessment:
           - Analyze inflation trends (CPI, PPI, PCE)
           - Evaluate economic strength (PMI, NFP)
           - Consider monetary policy outlook (FOMC)
           - Identify which sectors should perform best in this environment

        2. Sector-Level Analysis:
           - Determine sector overweight/underweight based on macro
           - Compare current sector exposure vs target allocation
           - Identify sectors requiring position changes

        3. Stock Selection Within Sectors:
           - Prioritize stocks in preferred sectors
           - Use sentiment data to rank within sectors
           - Consider existing positions (avoid unnecessary turnover)

        Provide response in format:

        MACRO ENVIRONMENT:
        - Current economic conditions
        - Key drivers
        - Sector implications

        SECTOR VIEWS:
        - Overweight sectors: [list with rationale]
        - Underweight sectors: [list with rationale]
        - Current vs Target exposure

        PORTFOLIO RECOMMENDATIONS:
        Positions to Long:
        1. [TICKER] (Sector: X)
           - Macro alignment: [explanation]
           - Sector view: [explanation]
           - Supporting sentiment: [relevant aspects]
           - Position size recommendation: [X%]

        Positions to Short:
        [Same format as above]

        TURNOVER ANALYSIS:
        - Summary of recommended changes
        - Rationale for maintaining existing positions
\end{Verbatim}

\subsection{Prompt for Ranking Agent in cross-sectional strategy}
\label{sec : prompt-ranking-agent-cross-sectional}
\begin{Verbatim}[breaklines=true, breakanywhere=true]
    You are a quantitative analyst specializing in sentiment-driven trading strategies. Your task is to analyze and rerank stocks for a long-short strategy based on sentiment data and macroeconomic context.

        Given the following inputs:
        [LIST OF STOCKS]
        - List of stocks that can be included in the portfolio
        - All stocks are assumed to start with the same score

        [SENTIMENT DATA]
        - List of stocks with sentiment-aspect pairs from news articles
        - Each pair contains: stock ticker, date, specific aspect (e.g., "management", "financial performance"), and sentiment score (-1, 0, 1)
        - Sample format: 2024-01-15|AAPL|[[management, 1],[revenue, -1]] (published_date|stock|aspect_sentiment_pairs)

        [MACRO DATA TRENDS]
        - Latest trend readings: CPI {value}, PPI {value}, PCE {value}, NFP {value}, PMI {value}
        
        [FOMC MINUTES SUMMARY]
        - Recent FOMC minutes summary: {key_points}

        Analyze how these macro trends might intesify or mitigate the sentiment towards different aspects. For example:
        - An increase in inflation might make cost-related sentiments more impactful
        - An increase in PMI readings might reduce supply chain concern impacts
        - Trends for Employment data might affect consumer demand sentiment importance

        For each stock, please:
        1. Evaluate each sentiment-aspect pair
        2. Adjust the importance of each aspect based on current macro conditions
        3. Assign a composite score that considers:
        - Sentiment scores
        - Macro-influenced aspect weights


        Then:
        1. Rank the stocks from highest to lowest composite scores
        2. Split into long candidates (positive scores) and short candidates (negative scores)
        3. Explain your reasoning for the each long and short picks
        4. Check thorugh the long and short candidates. If there are duplicates, review the composite scores and keep only the position with a higher composite score.
        
        Provide your response in this format:

        LONG CANDIDATES:
        1. [TICKER] - Score: [X]
        - Key aspects: [list most influential aspects]
        - Macro amplifiers: [which macro factors strengthened the case]

        2. [Continue for long picks...]

        SHORT CANDIDATES:
        [Same format as above]

        MACRO ANALYSIS:
        - Brief explanation of how macro conditions influenced the rankings
        - Which factors were most decisive

        Only provide factual analysis based on the data given.
\end{Verbatim}

\end{document}